\documentclass[11pt]{article}
\usepackage{amsmath}
\usepackage{amsfonts}
\usepackage{amssymb}
\usepackage[margin=1in]{geometry}
\usepackage{color}
\newtheorem{thm}{Theorem}
\newtheorem{conj}{Conjecture}
\usepackage[utf8]{inputenc}
\newcommand{\R}{\mathbb{R}}
\newcommand{\C}{\mathbb{C}}
\newcommand{\I}{\mathbf{I}^*}

\begin{document}

\begin{titlepage}
\hfill \\
\vspace*{15mm}
\begin{center}
{\Large \bf On Upper Bounds in Dimension Gaps of CFT's}

\vspace*{15mm}
{\large Tristan C. Collins$^1$, Daniel Jafferis$^2$, Cumrun Vafa$^2$, Kai Xu$^3$, Shing-Tung Yau$^{3,4,5}$}
\vspace*{8mm}

$^1$Department of Mathematics, Massachusetts Institute of Technology, Cambridge, MA 02138, USA\\
\vspace*{3mm}
$^2$Jefferson Physical Laboratory, Harvard University, Cambridge, MA 02138, USA\\
\vspace*{3mm}
$^3$Department of Mathematics, Harvard University, Cambridge, MA 02138, USA\\
\vspace*{3mm}
$^4$ Yau Mathematical Science Center, Tsinghua University, Beijing, 10084, China\\
\vspace*{3mm}
$^5$ Black Hole Initiative at Harvard University, 20 Garden Street, Cambridge, MA 02138, USA \\

\vspace*{0.7cm}

\end{center}

\begin{abstract}

We consider CFT's arising from branes probing singularities of internal manifolds.  We focus on holographic models with internal space including arbtirary Sasaki-Einstein manifolds coming from CY as well as arbitrary sphere quotients.  In all these cases we show that there is a universal upper bound (depending only on the spacetime dimension) for the conformal dimension of the first non-trivial spin 2 operator in the dual CFT and a minimal diameter (in AdS units) for the internal space of the holographic dual and conjecture it holds for all CFT's.

\end{abstract}
\end{titlepage}
\section{Introduction}
Conformal dimension of operator in CFT's has typically no large gaps near zero.  For example, the biggest known gap for all 2d CFT's is the very special `Monstrous Moonshine' construction which leads to the Monster group as its symmetry. In its left-right symmetric form, this theory has dimension gap of 4 for scalar operators and 2 for spin 2 operators.  In particular it has no relevant or marginal operators.
All known CFT's in $d>2$ have relevant or marginal operators.  Could it be that in any dimension there is a universal upper bound on the dimension of the first non-trivial operator in any CFT?

This question can be viewed in the context of holographic models which maps the CFT to a gravitational theory on $AdS^{d+1}\times M$ for some internal manifold $M$.  In such cases the existence of a large gap gets related to having no light particles.  In all the known examples for which we have analytic control, 
the operators that we get by this duality include some operators 
which are relevant or marginal. 
However in some proposed examples such as DGKT \cite{DeWolfe:2005uu}, whose validity is still being debated due to technical difficulties, even though there are low dimension spin 0 operators \cite{Cribiori:2021djm}, if consistent would imply the existence of arbitrarily large gaps for spin 2 KK modes (see also some related models \cite{Conlon:2021cjk}).

Even though the question of whether the first non-trivial CFT operator can have arbitrarily large conformal dimension is intrinsically interesting for a deeper understanding of CFT's, it also turns out to be deeply related to questions in the Swampland program.  Indeed generalizing the distance conjecture \cite{Ooguri:2006in} in the context of the Swampland program it was proposed in \cite{Lust:2019zwm} that if we have a family of AdS spaces with cosmological constant $\Lambda$, as $\Lambda\rightarrow 0$ we should get a tower of light states with masses which scale as 
$$m\sim |\Lambda|^{a}$$
for a number $a$ of order 1.
This conjecture in particular suggests that the limit of large AdS is accompanied by the limit of a decompactification of the internal space which lead to a tower of light states.  In other words the AdS scale cannot be separated from the scale of the internal geometry as in all known examples we have an extra space $M$ in the total space:
$AdS^{d+1}\times M$.  Indeed in all the examples in the literature where we have analytic control $a=1/2$.
This translates to the statement that the dimension of the tower of CFT operators %
$$\Delta \sim {m\over \Lambda^{1/2}}\sim m\ l_{AdS}\sim O(1)$$
In \cite{Lust:2019zwm} it was conjectured that $a={1/2}$ for all the supersymmetric cases which is consistent with the statement that we do not expect to obtain an arbitrarily large dimension gap.

In a sense it may appear surprising that we cannot choose the scale of the internal manifold $M$ independently of the cosmological constant.  For example in the case of Calabi-Yau compactifications where we obtain a Minkowski vacuum, the cosmological constant $\Lambda  =0$ and we can set the size of the Calabi-Yau independently of $\Lambda$.  In other words there is a scale separation in the usual Calabi-Yau compactifications, where we get a number of massless modes and the next excited mode can in principle be close to the Planck scale (depending on the moduli of the Calabi-Yau manifold).  This  motivates asking the following question:  Let us fix $\Lambda$ and ask if we can find a family of internal manifolds so that the states are arbitrarily massive?  This in particular translates to whether the spectrum of the Laplacian acting on the fields of our theory can have arbitrarily large first non-trivial eigenvalue leading to arbitrarily large mass gap.

One way to look for such a family is as follows:  Consider an internal manifold $M$ and assume we have a group action $G$ on it without fixed points leading to a smooth manifold $M/G$.  Clearly this will reduce the volume of $M$ by $|G|$. This results in a theory with the same $\Lambda$. If  $|G|\rightarrow \infty$ we can get the volume of the internal space to approach $0$.  Indeed there are $M$ and $G$ such that $G$ can act smoothly and $G$ can be arbitrarily large leading to volume of $M/G$ to be arbitrarily small.  So this may have led to our expectation that in this limit the eigenvalues of the Laplacian will be arbitrarily large and there would be a big gap in dimension of operators on the dual CFT.  As we will see in this paper, this expectation is not borne out.  In particular in all examples we have checked, even though the volume goes to zero, the diameter of $M/G$ surprisingly has a lower bound which leads to an upper bound for the first eigenvalue of the Laplacian.
For concreteness we consider the three cases that feature prominently in string theory where the internal manifolds include $S^3,S^5,S^7$ in detail and consider arbitrary quotients without fixed points.  We focus on the eigenvalues of the scalar Laplacian which sets a bound on the first spin 2 operator in the dual CFT. In particular we will not look for the lowest dimension operator because that is typically from a field with spin less than 2.  For spin 2, we find that no matter what $G$ we pick, there is an upper bound on how big the first eigenvalue can get.  In fact we find a sharp bound for smooth orbifolds of spheres.  Interestingly, even though one may have expected the biggest gap to come from some group action $G$ whose order goes to infinity, we find that the highest gap comes from finite order $G$ (and the icosahedral subgroup of $SU(2)$ is an important ingredient for such $G$).

One may have thought that the lower bound on the diameter of sphere quotients has arisen from the condition that $G$ has no fixed points.  However,
we can also consider $G$ actions with fixed points, even though these already lead to light modes coming from the singularities resulting from the fixed point action.  We show that even if you ignore the modes coming from singularities there is still an upper bound on the first non-trivial eigenvalue (and a lower bound on the diameter) of the sphere quotients (in any dimension), though in these cases it is harder to find this bound explicitly.

Instead of considering quotients to increase the minimum eigenvalue, one could try to look for a family of internal spaces for which the first eigenvalues can get arbitrarily large.  For example one could consider all Sasaki-Einstein manifolds of fixed curvature which appear as internal theories of holographic duals of branes probing CY singularities to see if the first eigenvalue can be made arbitrarily large.  We show that for arbitrary Sasaki-Einstein manifolds coming from regular CY 3-fold singularities there is an upper bound on the first non-trivial eigenvalue.  In particular this implies that for all ${\cal N}=1$ CFT's arising from type IIB D3 branes probing CY 3-folds with regular singularities we get a universal upper bound on conformal dimension of the first non-trivial operator.  Moreover we show the same holds for all the toric CY 3-folds with irregular singularities.  We also consider Sasaki-Einstein manifolds in all dimensions but in this case focus on those arising from Fermat type hypersurface CY singularities (the  Brieskorn-Pham families).  We find that in any dimension $n$ there is an $n$-dependent upper bound on the first non-trivial eigenvalue for this family.

For the case of D3 branes probing any Fermat type CY 3-fold singularities we find that for the chiral spin 2 operators the conformal dimension gap is bounded by $\Delta_1=202$.  This bound is saturated for D3 branes probing the CY 3-fold singularity given by the hypersurface  
$$z_0^{13}+z_1^{11}+z_2^{3}+z_3^2=0$$
Even though this is the sharp bound for largest dimension gap for chiral spin 2 operators for D3 branes probing any Fermat type Sasaki-Einstein manifold, there could be non-chiral operators with even lower conformal dimensions.

Based on these findings one may conjecture that {\it for any CFT in dimension} $d$, {\it the conformal dimension of the first non-trivial operator is bounded by a universal constant} $C_d$ {\it that depends only on} $d$.  It is natural to ask if $C_d$ is less than or bigger than $d$ for $d>2$.  In other words can there be a CFT in $d>2$ which has no relevant or marginal operators?  It is possible that the answer to this question is negative.  At any rate our results go in the direction of not expecting a large gap.  For example, if our conjecture is true, no 2d CFT no matter how large the central charge $c$, can have a gap bigger than a fixed number independent of $c$.  In this case it is natural to expect that the Monstrous Moonshine indeed saturates this upper bound (a recent work \cite{Lin:2021bcp} provides further support for such a picture). 

This physical conjecture motivates the following mathematical conjecture, motivated by the examples where branes probe point singularities of internal manifolds.  In such cases the internal manifold of the holographic dual becomes an Einstein manifold $M^n$ where
$$R_{ij}=(n-1)g_{ij}$$
(normalized so that the n-dimensional sphere has radius 1, which is relevant for finding the spectrum in AdS units).

\begin{conj}\label{conj}
The diameter of an $n$-dimensional Einstein manifold with positive Einstein constant $n-1$ is bounded from below and the first non-vanishing eigenvalue of scalar Laplacian bounded from above, where the bounds depend only on $n$.
\end{conj}

This conjecture holds for Kähler-Einstein manifold by the following argument: the Kähler class of a complex $m$-dimensional Kähler-Einstein manifold is $\frac{2\pi}{m-1}c_1(M)$, hence its volume is $(\frac{2\pi}{m-1})^mc_1(M)^m\geq (\frac{2\pi}{m-1})^m$ because $c_1(M)^m\in H^{2m}(M,\mathbb{Z})$ is a positive integer. Since the Ricci curvature is positive, by volume comparison theorem we see that the diameter has lower bound $(\frac{2\pi}{m-1})^{1/2}$. There is also analogous theorem for (nontrivial) gradient Ricci soliton \cite{ajm/1383923434}, which corresponds to internal manifolds with nontrivial dilaton background.  However, such an argument cannot generalize since there are examples of Einstein metrics on $S^5$ with Einstein constant $n-1$ and volume arbitrarily small \cite{Collins:2019}.

Of course we should note that not all the internal manifolds that arise in the context of holographic models are Einstein manifolds (see for example \cite{DeLuca:2021mcj}), as would be the case for spacetime filling branes wrapping also vanishing cycles of the internal manifold.  It would be interesting to extend the analysis in this paper to such more general classes.  Of course our conjecture on the CFT's, if correct, would imply a similar upper bound for the first non-trivial eigenvalue of the relevant Laplacians also in these cases.

The organization of this paper is as follows.  In section 2 we focus on the eigenvalues and diameter of spherical orbifolds and apply it to the cases of $S^3,S^5$ and $S^7$ (we also note a general bound for sphere quotients in arbitrary dimensions with or without fixed points).  In section 3 we find an explicit upper bound on the dimension of operators for Brieskorn-Pham (Fermat-type) family of Sasaki-Einstein manifolds in any dimension.  We also show that for an arbitrary Sasaki-Einstein manifold coming from singularities of regular CY 3-folds as well as all the toric irregular cases there is such on upper bound.  We also translate this to a conjecture for $d=2$, ${\cal N}=2$ SCFT's arising from sigma models on the corresponding singular CY's. 
Finally in section 4 we end with some concluding remarks.

\section{Dimension bounds for spherical quotients}
In this section we focus on CFT's for which the holographic dual involves a sphere in the internal geometry and we consider quotients of them in hopes of getting large gaps in dimension of operators.  In particular we aim to find a large first eigenvalue $\lambda_1>0$ for the scalar Laplacian. Taking a quotient seems the right thing to do to increase the eigenvalues because it would reduce the diameter of the space and one may naively think that by taking a large enough quotient we can get the diameter $D$ of the space as small as we wish.  We first focus on the case where the dimension of the sphere is odd.  This is because for even dimensional sphere any quotient (which will reside in $SO(2m+1)$) will have a fixed point and the fixed point will lead to massless or light modes which will automatically avoid having a large gap.

To begin with we focus on the first eigenvalue $\lambda_1$ and diameter $D$ of orbifolds $M=S^n/G$ for $n$ odd and $G$ a discrete subgroup of $SO(n+1)$ which acts without fixed points on $S^n$.
The cases of interest in physical applications are  $S^3,S^5$ and $S^7$.
The $S^3$ is part of the space for $AdS^3\times S^3\times K3$ for example.  The $S^5$ is part of the well known case of $AdS^5\times S^5$ and $S^7$ arises in dual of stack of $M2$ branes: $AdS^4\times S^7$. Here we always take $S^n$ to be the unit sphere, satisfying $R_{ij}=(n-1)g_{ij}$ and the cosmological constant $\Lambda=\frac{(n-1)(n-2)}{2}$. 

A generic choice for $G$ will break supersymmetry and makes the application of holography in the AdS/CFT context questionable (see e.g.\cite{Ooguri:2016pdq}). Nevertheless we will consider both supersymmetric and non-supersymmetric quotients as obtaining the bounds for eigenvalues and diameters are equally easy.  For supersymmetric quotients we would choose $G$ to lie in:
$$S^3\rightarrow G\subset SU(2)\times U(1)$$
$$S^5\rightarrow G\subset SU(3)$$
$$S^7\rightarrow G\subset SU(4)\ or\ Spin(7)$$
Spherical quotients in the context of holography for the $S^5$ case were first considered in \cite{Kachru:1998ys}, and their corresponding CFT duals were studied in \cite{Lawrence:1998ja}.

In the holographic context, to obtain all the dimensions we have to consider the spectrum of the Laplacian acting on various fields in the theory.  For simplicity we will only focus on the scalar Laplacian.  This is not only simpler to study, but it also is important physically for the following reason:  The scalar Laplacian eigensates, applied to the metric field, lead to spin 2 massive particles in $AdS^{d+1}$.  A bound on their spectrum thus puts a bound on the masses of the KK tower which is most relevant for scale separation.  In other words, there could even be massless/tachyonic modes in the gravity dual, leading to marginal/relevant operators in CFT.  But if the first spin 2 operator has a very high dimension, we can consider this as a scale separated theory in the sense that it decouples the higher spin states from light lower spin fields.  Thus to show that the CFT cannot have an arbitrarily large gap it suffices to put an upper bound on the first eigenvalue of scalar Laplacian of the internal manifold.

Let $\lambda$ denote any eigenvalue of scalar Laplacian.  This then translates to a spin 2 massive particle of (mass)$^2$ equal to $\lambda$.  For example for the sphere of unit radius the eigenvalues are given by 
$$\lambda= k(k+n-1),$$
where $k$ is the degree of the monomial. 
To translate this into the standard units in which the AdS curvature radius is set to 1, we must compare the radii of the internal sphere and AdS factors. 
For $AdS^5\times S^5$ and $AdS^3\times S^3 \times K3$, the radii are equal, while for $AdS^4\times S^7$, $R_{AdS} = \frac{1}{2} R_{S^7}$

These are then related to dimension $\Delta$ of spin 2 operators in the dual CFT by
$$\Delta(\Delta -d)=m^2=a\lambda=a k(k+n-1).$$
where $a = (R_{AdS}/R_{S^n})^2$.  For example for $AdS^5\times S^5$ where $d=4$ and $n=5$, $a=1$ and this leads to $\Delta =k+4$. For the $S^3$ and $S^7$ examples one obtains $\Delta =k+2,\Delta={k\over 2}+3$ respectively. 

The eigenfunctions correponding to holomorphic monomials (viewing the odd dimensional sphere as sitting in the complex space ${\bf C}^m$) preserve some amount of supersymmetry and lead to chiral operators on the CFT dual. The unitarity bound for spin 2 operators is $\Delta \geq Q + d$, where $Q$ is the $R$-charge \cite{Flato:1983te,Dobrev:1985qv,Minwalla:1997ka}. This is saturated for the graviton modes associated to holomorphic monomials on the internal manifold. 
For the sphere quotients without fixed points, eigenvalues are relatively straight-forward to study, as all we will need to do is to project out the monomials which are not invariant under $G$.   

The results we obtain which will be discussed later in this section are as follows:

For $S^3$:  $\lambda_1\in[3,168]$, $D\in(0.32,\pi]$;  The largest $\lambda_1$ arises for the icosahedral subgroup $G\subset SU(2)_{diag.}\subset SU(2)_L\times SU(2)_R$ and it is achieved by holomorphic eigenfunction.  This leads to $\Delta_1\in[3,14]$.

For $S^5$: $\lambda_1\in[5,32]$, $D\in(0.96,\pi]$; the largest $\lambda_1$ for $SU(3)$ quotients which preseve supersymmetry is 12 (instead of 32), and that cannot be achieved by holomorphic functions; This leads to $\Delta_1\in[5,8]$.

For $S^7$: $\lambda_1\in[7,40]$, $D\in(0.84,\pi]$; the largest $\lambda_1$ for $SU(4)$ quotients is also 40 and the group $G$ giving this is the icosahedral subgroup of $SU(2)$ which sits diagonally in the product of $SU(2)\times SU(2)\subset SU(4)$ extended by a $Z_2$ exchanging the two 2-dimensional subspaces, and can be achieved by holomorphic functions. This leads to $\Delta_1\in[3.5,5]$.

We should note that $\Delta_1$ denotes the lowest dimension non-trivial spin 2 operator coming from the modes on the sphere. In all these cases there are relevant operators ($\Delta <d$) with lower spin (or from other part of the geometry as in the $AdS^3\times S^3\times K3$).

\subsection{Some generalities}
First eigenvalue and diameter can be mutually controlled in both direction by Li-Yau estimate and Cheng's eigenvalue estimate: (they both stated the theorem for manifolds, but same proof works for orbifolds with quotient singularities as well):

\begin{thm} If $M$ is a compact $n$-dimensional Riemannian manifold with non-negative Ricci curvature, we have 
$$\pi^2\leq \lambda_1D^2<2n(n+4)$$
\end{thm}

But we will estimate them separately for accurate bounds. Here is the general strategy: $M$ has the form $S^n/G$ for a finite subgroup $G\subset O(n+1)$, and $G$ acts freely if we want $M$ to be a manifold. This imposes a very strong constraint on $M$ which gives a complete classification in all dimensions (roughly speaking, the building blocks are the cyclic groups and subgroups of $SU(2)$ classified by ADE Dynkin diagrams). In this case we can obtain exact bounds for both eigenvalue and diameter.

\subsection{Estimates of first eigenvalue}

We will begin with the estimate of the first eigenvalue. As we see in the next theorem, the eigenvalues of $M=S^n/G$ can be computed in an algebraic way:

\begin{thm}
The spectral decomposition of Laplacian on $S^n$ gives \[L^2(S^n)\simeq \bigoplus_{k=0}^\infty H_k\]
where $H_k$ is the space of harmonic polynomials on $\R^{n+1}$ homogeneous of degree $k$ and corresponds to eigenvalue $k(n+k-1)$ for $-\Delta$.
\end{thm}

This is a well known fact in spectral theory: we can extend eigenfunctions on $S^n$ to harmonic functions on $\R^{n+1}$ by multiplying a suitable power of the radius (here $0$ will be a removable singularity), and apply the Liouville theorem for harmonic functions on Euclidean spaces.

From this theorem we may obtain the following decomposition for sphere quotients, which reduce the problem to computing the harmonic polynomials fixed by $G$:
\begin{thm}
The spectral decomposition of Laplacian on $M=S^n/G$ gives \[L^2(S^n/G)\simeq \bigoplus_{k=0}^\infty H_k^G\]
where $H_k^G$ is the $G$-invariant subspace of $H_k$ and corresponds to eigenvalue $k(n+k-1)$ for $-\Delta$.
\end{thm}

Also we will use the following fact to produce harmonic polynomials from non-harmonic ones, which comes from an $sl_2$ action on polynomials generated by $\Delta$, $r^2$ and their commutator:
\begin{thm}
The natural multiplication map $$\C[r^2]\otimes H\longrightarrow \C[\R^{n+1}]$$
is an isomorphism, where $H$ is the space of harmonic polynomials on $\R^{n+1}$, $\C[\R^{n+1}]$ is the space of all polynomials on $\R^{n+1}$ and $\C[r^2]$ is the space of polynomials in $r^2$.
\end{thm}

As mentioned before, we have a complete classification assuming $M$ is a manifold, i.e. $G$ acts freely. Below we will follow the notations in \cite{Vinberg1993SpacesOC} and make use of the results presented there.  In particular we use the fact that there is a $G_\infty$ which is a limiting continuous group where all the discrete subgroups without fixed point embed in.

First let us consider the $5$ dimensional case as this is the easiest: $G$ can always be viewed as a subgroup of $G_\infty = (S^1)^3 \times A_3$. Hence we need only compute the fixed polynomials $\C[z,u,v,\Bar{z},\Bar{u},\Bar{v}]^{G_\infty} = \C[|z|^2 +|u|^2 +|v|^2, |zu|^2 + |uv|^2 + |vz|^2,|uvz|^2, (|u|^2-|v|^2)(|v|^2-|z|^2)(|z|^2-|u|^2)]$. The first nontrivial harmonic polynomial appears in degree $4$, which is $(|z|^4 + |u|^4 + |v|^4) -(|zu|^2 + |uv|^2 + |vz|^2)$. Hence it has first eigenvalue $4 \times (4 + 5-1) = 32$. A simple $G$ achieving this bound is given as follow: let $A=\text{diag}(\exp{\frac{2\pi i}{7}},\exp{\frac{4\pi i}{7}},\exp{\frac{8\pi i}{7}})\in U(3)\subset SO(6)$, $B=(321)\text{diag}(\exp{\frac{2\pi i}{3}},1,1)\in U(3)\subset SO(6)$, then $G=\langle A,B\rangle$ is a group which acts on $S^5$ freely and achieves this bound.

If we want the compactification to preserve certain supersymmetry, we would require $G'$ to be a subgroup of $SU(3)$, then $G_\infty'= (S^1)^2$, $$\C[z,u,v,\Bar{z},\Bar{u},\Bar{v}]^{G_\infty'}=\C[|z|^2,|u|^2,|v|^2,zuv,\Bar{z}\Bar{u}\Bar{v}]$$ and the first nontrivial harmonic polynomial appears in degree $2$ ($|u|^2-|v|^2$ and $|z|^2-|u|^2$),
hence the first eigenvalue is $2\times (2+5-1) = 12$. we can take the group generated by $A$ in the previous example. This bound is not achieved by holomorphic or antiholomorphic polynomials, the first holomorphic eigenvalue is $21$ with eigenfunction $zuv$.

Now we consider the $3$ dimensional case: $G$ can always be viewed as a subgroup of $G_\infty=S^1\times\I$ (here $\I$ is the isocahedral group, other discrete subgroups of $SU(2)$ also appear in the classification, but they are smaller than $\I$ and considering $\I$ gives rise to the optimal bound). Hence we need to compute the fixed polynomials $\C[z, w,\Bar{z},\Bar{w}]^{G_\infty}$. It’s easy to see that $\C[z, w,\Bar{z},\Bar{w}]^{S^1} = C[|z|^2 + |w|^2, |z|^2 -|w|^2, z\Bar{w},\Bar{z}w]$ and $\I$ fixes the first coordinate while acting on the other three as the standard three dimensional Euclidean symmetry. Let us denote these last three coordinates by $(x,y,t)$ and they are all harmonic. The first coordinate is an algebraic function of these three; in other words there are essentially three independent variables.

Hence we have reduced the question to finding the $\I$-invariant harmonic polynomials in its standard representation, which can be solved by its character theory. We can compute that the invariant polynomials have dimension $1,0,1,0,1,0,2$ in homogeneous degree $0$ to $6$, this implies that there is an invariant harmonic polynomial in degree $6$. The reason is as follows: We have a
trivial invariant polynomial $x^2+y^2+t^2$ together with its powers. Choose an invariant polynomial in degree 6 not proportional to $(x^2 + y^2 + t^2)^3$, it can be written as $f + (x^2 + y^2 + t^2)g$, with $f, g$ invariant, $f$ harmonic and $g$ has degree $4$. But the space of degree $4$ invariant polynomials is one-dimensional, hence $g = k(x^2+y^2+t^2)^2$ . By assumption, $f\neq 0$ and we find the invariant harmonic polynomial we want. Because $(x, y, t)$ are quadratic in $(z, w)$, $f$ has degree $12$ in $(z, w)$, and the first eigenvalue of the sphere quotient is $12 \times (12+3-1) = 168$. A simple $G$ achieving this bound would be $\mathbb{Z}_{13}\times \I\subset SU(2)\times SU(2)/ (-I,-I)\simeq SO(4)$. This cannot be achieved by holomorphic function, as it's only one dimensional (if it could be achieved by holomorphic function then it can also be achieved by antiholomorphic ones hence at least two dimensional), but if we only consider subgroups of $SU(2)$ it can indeed be achieved by a holomorphic one from the standard expression of du Val singularities (in this case the eigenspace is 12 dimensional, among which only one comes from holomorphic functions).

Next we consider the $7$ dimensional case. $G$ can always be viewed as a subgroup of $G_\infty = (S^1)^2 \times S_2 \times \I$ . Hence we need to compute the fixed polynomials $\C[z_1,\Bar{z_1},w_1,\Bar{w_1},z_2,\Bar{z_2},w_2,\Bar{w_2}]^{G_\infty}$. Adopting exactly the same approach as in the $3$ dimensional case, we write
\[
\begin{aligned}
&\C[z_1,\Bar{z_1},w_1,\Bar{w_1},z_2,\Bar{z_2},w_2,\Bar{w_2}]^{G_\infty}\\
&\quad =\C[|z_1|^2+|w_1|^2,|z_1|^2-|w_1|^2,z_1\Bar{w_1},\Bar{z_1}w_1,|z_2|^2+|w_2|^2,|z_2|^2-|w_2|^2,z_2\Bar{w_2},\Bar{z_2}w_2]^{S_2\times \I},
\end{aligned}
\]
and use the character theory of $S_2\times\I$ to get the first invariant harmonic. It happens in degree $4$, hence the first eigenvalue is $4\times(4+7-1) = 40$. In fact, we can construct the eigenfunction by hand: $x_1x_2 +y_1y_2 +t_1t_2$ satisfies the requirement (with similar notation in $3$ dimensional case). We can produce such $G$ by modifying the previous case: we let $\I$ acts diagonally on $\R^8\simeq \R^4\times \R^4$ and $\mathbb{Z}_{13}$ acts on two copies with different characters. Actually this can still be achieved by subgroups of $SU(4)$ (by diagonal $E_8$ action and swapping with a $-1$ twisting).

Now we have shown that the optimal upper bound of first eigenvalue of sphere quotients in dimension $3,5,7$ are $168,32,40$ respectively. This upper bound can be achieved by many different sufficiently complicated finite groups.





 \subsection{Estimates of diameter}

 Now we proceed to the estimation of the diameter, which involves more complicated geometric calculations. In the manifold case, as before we can compute the exact bound for diameter $D$ by the classification in \cite{Vinberg1993SpacesOC}. 

 In dimension 5, we can take a sequence of $G\subset O(6)$ converging to $(S^1)^3\times A_3$, and the limiting quotient space is a equilateral triangle quotient by its cyclic symmetric group, hence the diameter is explicitly calculable to be $\arccos{(3^{-\frac{1}{2}})}\approx 0.96$. 

 If we want $G$ to be a subgroup of $SU(3)\subset O(6)$, we can only take a sequence converging to $(S^1)^2\times A_3$ and then the diameter is similarly computed to be $\frac{\pi}{3}\approx 1.05$.

 In dimension 3, the largest limiting group is $S^1\times \I$, and the quotient would be $S^2/\I$. The diameter is the distance from the vertices to the face center of an icosahedron (considered as a cell structure on $S^2\simeq S^3/S^1$), and can be computed to be $\sin(\frac{1}{2}\arccos\sqrt{\frac{1+2\cos\theta}{3}})\approx 0.32$ where $\theta=2\arcsin\frac{1}{2\sin(\frac{2\pi}{5})}$ is the length of an edge of the icosahedron.

 The dimension 7 case is a doubled version of the previous one: here we need to consider the limiting group $(S^1)^2\times S_2\times \I$ acting on two copies of the four dimensional representation, and a detailed examination proves that the diameter is $\arccos(\frac{1}{\sqrt{2}}\cos(\frac{1}{2}\arccos\sqrt{\frac{1+2\cos\theta}{3}}))\approx 0.84$.


\subsection{Estimates for spherical orbifolds with singularities}

For orbifolds with singularities we can still prove boundedness but the bound we get is rather loose. Hence we will only give a sketch of the arguments below (the reader is referred to \cite{article} for a stronger result with complete proof).

A first observation is that, if we decompose a large Hilbert space into orthogonal subspaces $V=V_1\oplus V_2$, any two unit vectors $e_i\in V_i$ have equal distance $\sqrt{2}$. Hence if $G$ action preserves $V_i$, any pair $g_1e_1,g_2e_2$ still have equal distance $\sqrt{2}$, which means that in the quotient space $V/G$, distance between $e_i$ is $\sqrt{2}$, hence the unit sphere (with Euclidean distance inherited from $V$) has diameter at least $\sqrt{2}$.

This argument can actually be pushed further to more general group actions. We define a representation to be primitive if it cannot be decomposed nontrivially into several subspaces which are permutted by the group action, i.e. we cannot write $V=\oplus V_i$ such that for any $g\in G$ and $V_i$, $gV_i=V_j$ for some $j$ in a nontrivial way. For example, only primitive subgroups of $SU(2)$ are $E_6,E_7,E_8$, although the fundamental representation is also irreducible when restricted to $D_n$. If $G\subset O(n+1)$ is not primitive, the problem of constructing eigenfunction can be reduced to the (strictly smaller) subgroup of $G$ which maps each subspace into itself, and the diameter is bounded in an easier way: Suppose $V_1$ is mapped to $V_2$ by some element in $G$. take $e_1\in V_1, e_2\in V_2$, then $\mathrm{dist}_{S(V)}(\frac{e_1+e_2}{\sqrt{2}},e_1)=\mathrm{dist}_{S(V)}(\frac{e_1+e_2}{\sqrt{2}},e_2)=\frac{\pi}{4}$ (where ${S(V)}$ means unit sphere in $V$ and we are considering the spherical distance). This implies that, if we begin with an arbitrary point $p$ in $V_1$, any of its image under $G$ must lie in either $V_1$ or $V_2$, and hence has distance at least $\frac{\pi}{4}$ with $\frac{e_1+e_2}{\sqrt{2}}$, therefore the diameter of $S(V)/G$ is at least $\frac{\pi}{4}$.\footnote{This also proves the boundedness of first eigenvalue by Li-Yau estimate in this context.}

Hence we only need to consider the primitive case. It's a classical result \cite{blichfeldt1917finite} that such groups modulo center have order bounded by $2(n+1)!5^{n\theta(n+1)+2}$ (here $\theta(n+1)$ is the number of primes less than or equal to $n+1$, this bound is very far from optimal). This bound of group order will allow us to bound the first invariant polynomial (and hence first invariant harmonic polynomial by the previous theorem 5) by a classical theorem of Noether in invariant theory.

To estimate the diameter we can similarly reduce to primitive $G$ where we can bound its order, hence the volume of the quotient and therefore its diameter. Just as before, as the bound for order is extremely loose, the diameter bound we get is far from optimal, and has order $10^{-(n-1)}$ for dimension $n$ orbifold quotients.

\section{Dimension Bounds for Sasaki-Einstein Manifolds}

\subsection{Brieskorn-Pham (Fermat-type) family}
In this section we consider a generalization of previous estimates to a larger family of Einstein manifolds coming from (5 dimensional) links $L(a)$ of Brieskorn-Pham singularities $Z(a)\subset\C^4$ defined by the base of the cone for the singularities of the type
\[f(z)=\sum_{i=0}^3 z_i^{a_i}=0\]

The corresponding holographic setup arises in type IIB where we have $N$ D3 branes probing the singularity $Z(a)$ of a Calabi-Yau 3-fold, leading to the $L(a)$ as the internal manifold of the holographic dual.
The link $L(a)$ has a natural Sasakian structure (which is an odd dimensional cousin of Kähler structure), and it's expected that its link admits an Sasaki-Einstein metric whenever a unitarity bound is satisfied, in the sense that the minimal weight of a holomorphic function is strictly greater than one, i.e. $0<\sum\frac{1}{a_i}-1<3\min_j\frac{1}{a_j}$ (see, for example, \cite[Note 36]{boyer2005einstein} and \cite[Theorem 8.1]{Collins:2019} ).  These bounds are related to unitarity bounds on the dual ${\cal N}=1$ SCFT \cite{Gauntlett:2006vf}.  Also, the lower bound $(\sum {1\over a_i}>1$) is the same condition as the singularity appearing at finite distance in moduli space \cite{Gukov:1999ya}. From the string worldsheet this condition has the interpretation that the corresponding string theory has a worldsheet CFT including an ${\cal N}=2$ SCFT with superpotential $W=Z$, leading to central charge
$$ {\hat c}=\sum(1-{2\over a_i})<2$$
consistent with the fact that the deficit to $\hat c=3$ to give a 3-fold CFT is provided by a linear dilaton background which necessarily has $\delta {\hat c}>1$ \cite{Ooguri:1995wj}. Such Sasakian structures arising from algebraic singularities are called quasi-regular and in this case we have a one to one correspondence among following three objects: conical Calabi-Yau metric on the singularity $Z(a)$, Sasaki-Einstein metric on the link $L(a)$ and Kähler-Einstein metric with positive curvature on the projectivization $X(a)$ (i.e. the quotient of the link by natural $S^1$ action $L(a)/S^1$). 

For such manifolds it is difficult to compute the eigenfunctions and eigenvalues explicitly as we have done for the sphere, with the exception of holomorphic polynomials in $z_i$. For these holomorphic polynomials, we know that they correspond to holomorphic sections of tautological line bundles on $X(a)=L(a)/S^1$, hence the Laplacian on the base of this $S^1$ fibration gives $0$, and the Laplacian eigenvalue on the total space $L(a)$ can be computed by restricting to the fiber, i.e. counting the degree. In contrast, other harmonic sections on $X(a)$ would depend on the details of its Kähler-Einstein metric which are much more complicated. 

More precisely, if we assign degree $\frac{1}{a_i}$ to $z_i$, $Z(a)$ is defined by a degree one polynomial $f$, and its holomorphic volume form $$\text{Res}_{Z(a)}\frac{\prod dz_i}{f(z)}$$ has degree $\sum \frac{1}{a_i}-1$. As we wish to have the same cosmological constant as with the sphere, we may assign degree $\frac{1}{a_i}\times\frac{3}{\sum \frac{1}{a_i}-1}$ to $z_i$ so that the holomorphic volume form would have same degree as before, and we may check the relation between degree and eigenvalue also remains the same.
Note that this is simply the statement that the holomorphic 3-form should have degree 3 and the bounds follow from the usual bounds on the R-charge of the chiral operators.

Now let us try to maximize the degree of first holomorphic function, which is equivalent to minimizing $F(\mathbf{a})=F(a_0,a_1,a_2,a_3)=a_0(\sum\frac{1}{a_i}-1)$ (where we have assumed $a_0=\max{a_i}$)
\begin{itemize}

\item If $F(a_0,a_1,a_2,a_3)<F(13,11,3,2)$ we must have $\frac{1}{a_1a_2a_3}<F(13,11,3,2)=\frac{1}{66}$ as 
$\sum\frac{1}{a_i}-1\geq \frac{1}{a_0a_1a_2a_3}$

Clearly $a_3>1$, as otherwise we will have $F>1$ and $a_3<4$ by contradiction.

\item If $a_3=3$, we have $3\leq a_2\leq 4$ by the constraint, and then $a_1\leq 6$, which contradicts the fact $a_1a_2a_3>66$.

\item If $a_3=2$, we have $2\leq a_2\leq 5$ by the constraint. If $a_2=2$, $F>1$, a contradiction; if $a_2=3$, we have $ a_1\leq 11$ by the constraint, which contradicts the fact $a_1a_2a_3>66$. If $4\leq a_2\leq 5$, we have $ a_1\geq 7$ by the fact $a_1a_2a_3>66$, which contradicts the constraint.

\end{itemize}

Hence we see that $\min F=F(13,11,3,2)=1/66$. Topologically this link is also an $S^5$ (with an exotic Einstein metric) \cite{boyer2005einstein}. From this we can see that the optimal bound for degree of holomorphic functions is 198 corresponding to $\Delta_1=202$, but this does not exclude the possibility of finding smaller eigenvalues which do not come from holomorphic or anti-holomorphic functions, corresponding to non-chiral operators. In comparison, the sphere quotients by $SU(3)$ subgroups have first eigenfunction in degree $3$ (corresponding to $\Delta_1= 7$) and eigenvalue 21, so it is reasonable to expect indeed there are smaller non-holomorphic eigenfunctions.

The combinatorial analysis above above actually applies to any dimension, which gives a finite algorithm and determines a finite upper bound for the first holomorphic eigenfunction. More precisely, we only need to consider $a_i$ which satisfy $\sum_{i>0}\frac{1}{a_i}<1$ by induction on dimension, since otherwise $\mathbf{b}=(b_0=a_1,\cdots,b_{n-1}=a_n)$ gives a singularity which is one dimensional lower and satisfies $F(\mathbf{b})<F(\mathbf{a})$. By induction on dimension $F(\mathbf{b})$ is bounded below and thus so is $F(\mathbf{a})$. The constraint $\sum_{i>0}\frac{1}{a_i}<1$ gives us a finite set, as none of the parameters can grow unboundedly, hence gives a positive minimum of $F$ and finite maximum of first holomorphic eigenfunction. We can get the exact bound by the algorithm above, for $n=4$ ($7$ dimensional link of Calabi-Yau fourfold singularity), the maximum is achieved by $(a_i)=(85,83,7,3,2)$, which gives degree $2\times 3\times 7\times 83\times 4=13944$. This would suggest that for the M2 branes probing the corresponding Fermat type CY4-fold singularity the gap for chiral spin 2 operator is bounded by $\Delta_1=3+{13944\over 2}=6975$ which is saturated in this example.

Another proposal in the literature \cite{Polchinski:2009ch} for obtaining large gaps is to consider M2 branes probing singularities in elliptic fibered CY4-folds. However the cones over Sasaki-Einstein manifolds cannot contain compact holomorphic subvarieties, and thus cannot be elliptically fibered.\footnote{The examples in \cite{Polchinski:2009ch} do not have a $U(1)_R$ symmetry, and thus are not conical.} Therefore the geometry near the singularity in such manifolds must be very different, and the diameter of the CY manifold far away will likely not provide a good guide to the near horizon geometry.  

In general, the maximum is achieved by the $a_n=2,a_{k}=\prod_{l>k}a_l+1\,\,(k>1), a_1=2\prod_{l>1}a_l-1,a_0=2\prod_{l>1}a_l+1$, which can be argued as follows: we always have $F\geq \prod_{i>0}\frac{1}{a_i}$ as their ratio is a positive integer, and we only need to minimize $\prod_{i>0}\frac{1}{a_i}$ in the finite set of $(a_i)$ we defined above. Note that when $F$ is small, $\sum\frac{1}{a_i}$ is necessarily very close to $1$, which can be viewed as restricting $\frac{1}{a_i}$ to the neighbourhood of a hypersurface, and we want to minimize the product $\prod_{i>0}\frac{1}{a_i}$, which is a convex function, hence we only need to consider the extreme cases, i.e. taking boundary of all constraints. There is only a finite number of choices, and we wish to minimize the geometric mean of $\frac{1}{a_i}$ for $i>0$ and maximize $\frac{1}{a_0}$, hence we take the left boundary point for $a_{0},a_n$ and right boundary point for $a_{i}(0<i<n)$, which gives the sequence we constructed. By $\cite{boyer2005einstein}$ this link actually admits a Sasaki-Einstein metric.

\subsection{General Sasaki-Einstein manifolds}

It's natural to ask what happens for links coming from more general algebraic singularities.  As we will show here, at least in the case of CY 3-folds indeed they follow a similar pattern as the Brieskorn-Pham family already discussed.

First, we may observe that in the expression of eigenvalue, only the $U(1)$ degree of $z_i$ and the monomials of $z_i$ occurring in $f$ appear, this is insensitive to the deformation of the singularity. For example, any homogeneous polynomial (with isolated singularity at the origin) determine same holomorphic eigenvalue with the Euler one (which is the case where all $a_i$ are equal). Hence we only need to study the Calabi-Yau singularities up to deformation equivalence, and under very mild assumption complete classification is known in dimension $3$. (Complete intersection singularities classified in \cite{Xie:2015rpa,Chen:2016bzh} and quotient singularities classified in \cite{Chen:2017wkw}). Just like the classification of CY2 singularity\footnote{There is one little difference here: in the CY2 case there are no deformation of singularities but in the CY3 case we do have nontrivial deformation parameters which are irrelevant to our discussion} where there are two infinite families ($A_n,D_n$) and a finite family ($E_6,E_7,E_8$), in the three dimensional case there are also a finite number of (infinite or finite) families exhausting all possible deformation types. The number of families is huge (even for hypersurface singularities there are 19 types each classifies into dozens of families), but the finiteness already implies that there is a finite upper bound for first holomorphic eigenvalue. In each family (infinite or finite) we can write down the first holomorphic eigenvalue explicitly as a rational expression of certain parameters and directly see an exact upper bound for this family, since there are only finitely many families we can just take their maximum and get a finite upper bound. 

Although the boundedness for each infinite family can be proven case by case directly, we now show that for the hypersurface \cite{Xie:2015rpa} case it follows from a simple argument.  For each family we have a uniform way to assign (positive) weights $q_i$ to $z_i$, so that $\sum q_i>1$ in this family. For an infinite family labeled by $k$, as $k$ goes to infinity, one of the $q_l$'s goes to $0$, and we see that the sum of rest of $q_j$ is still at least one. This implies that, $\sum q_i-1$ is at least $q_l$ and after normalization (so that the weight of volume form $\sum q_i-1$ is replaced by $n=3$), the weight of $z_l$ is at most $3$. Here we see an explicit upper bound $3$ of weight for all infinite families (and hence an explicit bound $21$ of eigenvalue), which is very small compared with the degree bound $198$ for type I (which of course lie in a finite family in type I). Hence we see that in general the maximizer should still lie in a finite family. 


In the previous section we found the optimal bound for Brieskorn-Pham (Fermat-type) singularities (which is type I in their classification) by a convexity argument.
Even though for arbitrary regular CY3-fold singularities the argument above proves boundedness of the first eigenvalue and gives a finite algorithm determining the optimal bound, it is cumbersome to compute it explicitly in this way due to the huge number of families in the classification. 

So far we have only focused our discussions on regular and quasi-regular Sasaki-Einstein manifolds, which corresponds to $U(1)$ bundles over Kähler-Einstein manifolds (without or with quotient singularity). There is another type named irregular Sasaki-Einstein manifolds, where the Reeb vector field does not integrate to a $U(1)$ action and has eigenfunctions with irrational eigenvalues. These correspond to dual ${\cal N}=1$ SCFT's in 4d where a-maximization leads to irrational R-charges. A large class of examples is obtained from the links of toric Gorenstein singularity\footnote{For Sasaki-Einstein manifolds this can be generalized to $\mathbb{Q}$-Gorenstein singularities, but then there is no Killing spinor, i.e. supersymmetry is lost.} \cite{sparks2010sasaki} (including examples $Y^{p,q}$ \cite{martelli2006toric} and $L^{a,b,c}$ \cite{martelli2005toric} in dimension $5$). In this case we have a uniform upper bound on the degree of first holomorphic eigenvalue given by complex dimension $m$ of the toric singularity $X$, achieved by the holomorphic function $\frac{\Omega_X}{\Omega_T}$, where $\Omega_X$ is the canonical volume form on the singularity (whose existence is equivalent to the Gorenstein property) and $\Omega_T$ is the invariant volume form on the torus $T\subset X$. $\Omega_X\wedge\Bar{\Omega}_X$ is the (real) volume form on the Kahler cone, hence has degree $2m$, thus $\Omega_X$ has degree $m$. $\Omega_T$ is invariant by definition, hence has degree $0$, and  $\frac{\Omega_X}{\Omega_T}$ has degree $m$. In particular, for $5$ dimensional Sasaki-Einstein manifolds $Y^{p,q}$ and $L^{a,b,c}$, there is always a degree $3$ holomorphic eigenfunction.

In the context of Sasaki geometry, the existence of a Sasaki-Einstein metric is equivalent to the algebraic notion of K-stability \cite{Collins:2019, Collins2:2018}; see also \cite{collins2016k} for physical aspects.  In particular, it may be possible to use the algebraic machinery of K-stability to prove that the lowest weight of a holomorphic function on a Calabi-Yau cone must be bounded from above.  Since any such holomorphic function is naturally an eigenfunction of the Laplacian on the link of the cone, with eigenvalue depending on the weight, this would yield an algebraic proof of Conjecture~\ref{conj} in the setting of Sasaki geometry.  It is worth noting, however, that the standard construction of the Rees algebra associated to a homogeneous holomorphic function yields only {\em lower} bounds for the weights \cite{Collins2:2018}; indeed, this constructions recovers precisely the unitarity bound of \cite{Gauntlett:2006vf}.  On the other hand, the examples of the Brieskorn-Pham singularities discussed earlier show that Conjecture~\ref{conj} holds even when the cone does not admit a Calabi-Yau metric (since many of the Brieskorn-Pham links are not K-stable, see, e.g. \cite{Collins:2019,boyer2005einstein}).  Indeed, it is reasonable to expect that if Conjecture~\ref{conj} holds, then it may also hold for metrics with any definite, positive lower bound for the Ricci tensor.    

\subsection{A conjecture for ${\cal N}=(2,2)$ SCFT's in d=2}
The above general conjecture for the bound on the holomorphic eigenvalues of general Sasaki-Einstein manifolds, combined with physical arguments, follows from a conjecture for $d=2$ SCFT's with ${\cal N}=2$
supersymmetry.  This follows from the behaviour of the expected sigma model for the corresponding CY singularity.  Namely we expect that point singularities of CY $n$-folds to give rise to ${\cal N}=2$ SCFT sigma models which is essentially a tensor product of linear dilaton background with an ${\cal N}=2$ SCFT \cite{Ooguri:1995wj}.  The bounds for the corresponding eigenvalues of the Laplacian lead to the following conjecture: 

{\it Consider an arbitrary ${\cal N}=(2,2)$ SCFT in $d=2$ such that $\hat c< k$ where $k\in {\bf Z}$ and let $q$ be the R-charge of the lighest chiral operator.  Then}
$${q \over k-{\hat c}}\leq C_k$$
{\it where $C_k$ is a finite constant depending only on $k$.  Moreover this bound is achieved by an SCFT}. In other words if we have a family of SCFT's for which ${\hat c}$ approaches an integer, then the lighest chiral operator should have a light dimension not much bigger than how close $\hat c$ is to that integer. For example if $k=1$ this corresponds to minimal model SCFT's and the equality is achieved for the $E_8$ minimal model with $C_k=3$. 

Given what we have said before this conjecture will already give us the bound for the Sasaki-Einstein manifolds coming from CY $k+1$
folds.  Of course if true this conjecture is more general as it includes other possibly non-geometric stringy backgrounds. 

\section{Concluding Remarks}
In this paper we have focuced on holographic models arising from branes probing internal singularities of CY manifolds and have shown that for a large class of models there is an upper bound for the first eignevalue of the scalar Laplacian of the internal (Sasaki-)Einstein manifold.  This leads to the statement that the conformal dimension of the first non-trivial spin 2 operator is bounded and cannot be arbitrarily large.

Based on these examples it is natural to conjecture that there is a universal bound on the conformal dimension of the first non-trivial operator for any CFT which depends only on the dimension of spacetime.  It would be important to try to check whether this conjecture is true by considering even wider class of examples (including branes wrapping vanishing cycles of internal manifolds).  It would also be important to try to find an argument for such a bound using bootstrap techniques or to construct counter-examples.

\subsection*{Acknowledgments}
 We would like to thank Igor Klebanov for a helpful discussion. The work of TC is supported in part by an Alfred P. Sloan fellowship and NSF CAREER grant DMS-1944952. The work of DJ is supported in part by the DOE grant  DE-SC0007870. The work of CV is supported in part
by a grant from the Simons Foundation (602883, CV) and by the NSF grant PHY-2013858. The work of STY is supported in part by a grant from the Simons Foundation (385581) and by a grant from John Templeton Foundation (61497).

\bibliographystyle{unsrt}
\bibliography{ref.bib}

\end{document}